\documentclass{emulateapj}
\usepackage{natbib} 
\usepackage{placeins}       
\usepackage{rotate,rotating}
\usepackage{epsfig,graphicx}
\usepackage{float}
\usepackage[pass]{geometry}
\def\lap{\lower.5ex\hbox{$\; \buildrel < \over \sim \;$}}
\def\gap{\lower.5ex\hbox{$\; \buildrel > \over \sim \;$}}

\def\ergcm2s{${\rm erg\ cm^{-2}\ s^{-1}}$}

\def\ergscm2s{${\rm erg\ cm^{-2}\  s^{-1}}$}

\def\cm-2{${\rm cm^{-2}}$}

\begin{document}
\title{Reducing and Analyzing the PHAT Survey with the Cloud}

\author{Benjamin F. Williams\altaffilmark{1},
Knut Olsen\altaffilmark{2},
Rubab Khan\altaffilmark{1},
Daniel Pirone\altaffilmark{3},
Keith Rosema\altaffilmark{4}
}
\altaffiltext{1}{Department of Astronomy, Box 351580, University of Washington, 
Seattle, WA 98195; ben@astro.washington.edu; rubab@uw.edu}
\altaffiltext{2}{National Optical Astronomy Observatories, 950 Cherry St., Tucson, AZ; olsen@noao.edu}
\altaffiltext{3}{Forehead Consulting, 2326 31st Ave South, Seattle, WA 98144; dpirone@forehead.com}
\altaffiltext{4}{Random Walk Group LLC, 501 Columbia Street NW, Box H, Olympia WA, 98501; rosema@randomwalkgroup.com}

\begin{abstract}

We discuss the technical challenges we faced and the techniques we used to overcome them when reducing the PHAT photometric data set on the Amazon Elastic Compute Cloud (EC2).  We first describe the architecture of our photometry pipeline, which we found particularly efficient for reducing the data in multiple ways for different purposes.  We then describe the features of EC2 that make this architecture both efficient to use and challenging to implement. We describe the techniques we adopted to process our data, and suggest ways these techniques may be improved for those interested in trying such reductions in the future.  Finally, we summarize the output photometry data products, which are now hosted publicly in two places in two formats.  They are in simple fits tables in the high-level science products on MAST, and on a queryable database available through the NOAO Data Lab.

\end{abstract}

\section{Introduction}

Large catalogs are an important component of observational astronomy.
Entire research programs that only analyze public catalogs are
becoming common \citep[e.g.,][]{yang2007,ivezic2007,shen2011,kleinman2013,tempel2014,salim2016,tian2017,bobylev2017}.  The production of these catalogs,
and the interfaces by which they are served to the public will be of
significant concern in the coming decade, especially for projects like
LSST and missions like WFIRST, which will be cataloging large numbers
of measurements of billions of sources \citep{ivezic2008,spergel2013,williams2015}.  One option for
handling the computational challenges surrounding producing and
analyzing these catalogs is using commercial cloud compute
services such as Amazon's Elastic Computing Cloud (EC2).

There are an increasing number of computer use cases in astronomy
research that require a large amount of computing power for small
periods of time.  For such projects, it is not efficient to purchase
large compute clusters or a large number of computers.  Furthermore,
many institutions do not allow users to have root access to their
machines.  However, commercial services that provide computing at an
hourly rate allow researchers to obtain access to a nearly unlimited
amount of computing power.  If the amount of time they need is
relatively short, the cost can be much cheaper than purchasing
machines.  Furthermore, they can have complete control of the
operating system as root users, simplifying software installation and
allowing streamlined memory use.

With a cloud-based system, one can hire as many processors as
necessary only for the time needed to do the work.  Thus, if a
processing job that takes a week on a single machine can be
distributed among hundreds of machines, it can be done in a few
minutes for a few dollars, without the need to purchase a large
compute cluster.

This capability can be quite cost-effective for astronomers, as
observational data sets can be quite large, but many programs may only
need to reduce one very large data set each year.  This reduction may
require a large amount of computing, but once it is done for the year,
the need for a large amount of computing power is gone, making
purchasing computers highly inefficient in such cases.

Another difficulty surrounding working with large data sets is sharing
the output catalogs with the community.  Once the photometric
measurements are complete, they should be served in a format that is
useful for the community.  The Sloan Digital Sky Survey
\citep[SDSS][]{york2000,aihara2011,abolfathi2017} set the standard for
one solution to this problem, which is to serve the catalogs using the
SkyServer \citep{Szalay2001} web-based database interface.

SDSS SkyServer serves a Microsoft SQL Server database with a flexible
web-based search form that is now typical for many catalog services,
such as Vizier \citep{ochsenbein2000} and the SAO/NASA Astrophysics
Data System \citep[ADS][]{kurtz2000}.  For SDSS catalog access, its
most powerful features are the ability to issue direct SQL queries to
the server both in real time and in batch mode, and to store results
in personal database tables.  This approach allows users to formulate
scientific questions as queries with all the necessary complexity,
creating subsamples of the catalog to be downloaded for further
analysis.

For future large surveys, such as the 10-year survey to be undertaken
by the Large Synoptic Survey Telescope \citep[LSST][]{lsst2009}, the
data volume requires new database technology development, and even
catalog subsamples will often be too large to download efficiently.
LSST will extend the approach used by SDSS, developing a new database
architecture to be able to scale to LSST's data requirements
\citep{wang2011}, and allowing complex analysis to run on servers
close to the catalog and image data itself (through the LSST Science
Platform), thus avoiding massive downloads of data.  Indeed, a number
of projects have begun implementing such science platforms for use on
existing large datasets, including CANFAR \citep{Gaudet2010},
SciServer (http://sciserver.org), and the NOAO Data Lab \citep{Fitzpatrick2014}.  With large volumes of data becoming common for many
smaller programs, the need for such platforms to provide efficient
data interfaces is growing, in particular because small programs
rarely have the resources to develop such interfaces for themselves.


One of the first large survey programs to be reduced using commercial
cloud resources was the PHAT survey \citep[][hereafter,
  W14]{williams2014}. Herein we will describe the challenges faced and
techniques used in reducing the PHAT data using a cloud service.  We
used Amazon Web Services (AWS), so we discuss AWS applications in
detail, but but other cloud service providers (such as Google Cloud
Services, Microsoft Azure, Oracle Cloud etc.) also offer similar
functionality.  Section 2 gives an overview to the technical aspects
of EC2.  Section 3 describes the PHAT pipeline architecture, which
works well on such cloud computing platforms.  Section 4 discusses the
importance of monitoring systems for keeping track of pipeline jobs
and handling problems.  Section 5 describes our photometry database
schema, and section 6 provides an example of how to use the database.
Finally, we conclude with a summary of the paper in Section 7 and a
brief discussion of newer cloud computing features in the Appendix.

\section{Amazon Cloud Computing and Storage}

Amazon offers on-demand computing resources primarily through the AWS
Elastic Compute Cloud (EC2). As with other cloud computing services,
this service has a virtually unlimited amount of compute power
available at any given time, but one must access and use the resource
carefully to make it cost-effective.  In this section, we describe the
various aspects of EC2, and how they work together to provide a
powerful computing resource for the community.

\subsection{Instances}

Computer clouds are large numbers of individual computing cores.
Amazon calls each of these cores ``instances,'' which are compute
nodes that are assigned to user accounts upon request.  Launching
these instances requires an Amazon Web Services account, as well as an
Amazon Machine Image (AMI).  Users are then charged for the amount of
time each requested instance is running.

Instances have multiple pricing options, which range over an order of
magnitude.  The most expensive is "on-demand."  If an instance with
this pricing option is requested it instantly starts booting up, and
it will stay operational until the user explicitly terminates it.
These instances have very high reliability, but are much more
expensive than simply buying a computer, if they are going to be up
for more than a few months per year.  However, they may still be
  of great use to those who either wish to have a reliable instance in
  the cloud for quick and free access to storage, high bandwidth
  communication to other instances, or rapid and reliable
  availability.

The next option is "reserved."   These instances are rented for
  periods of years as a contract.  Once they are reserved, they stay
  up for the full year, whether they are in use or not.  This is the
best option for instances you know you will need to be up for many
months, since they are typically cheaper than on-demand by a factor of
a few if both are being constantly used for months at a time.

\begin{figure*}
\begin{center}
\includegraphics[width=6.0in]{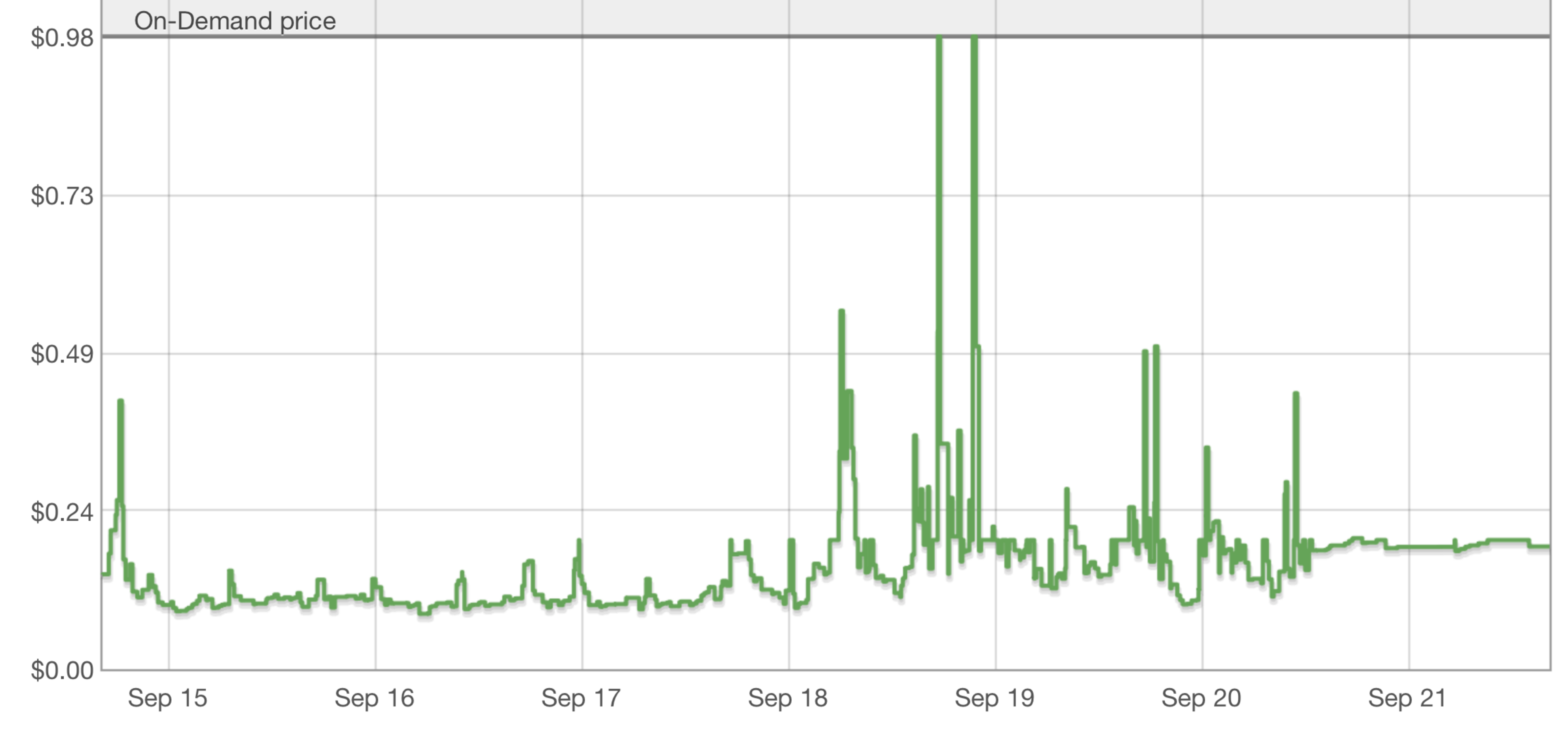}
\end{center}
\caption{Example of the pricing of a spot instance over time.  Occasionally the rate can meet or exceed the on-demand price, but it is usually much lower.  In this example, a bid of \$0.24 was safe for a few days from Sep. 15 to Sep. 18.}
\label{spot_price_example}
\end{figure*}

Finally, there is the cheapest option, which is called "spot."  The
price of these instances is dynamic, and changes by the minute.
Typically, the price is less than 20\% of the on-demand price.  One
specifies a maximum price for these instances, and if the price goes
above that maximum, the instance is automatically terminated without
warning.  When the price is below the bid, the user pays the price,
not what they bid, so that if the price is one cent per hour and the
bid is one dollar per hour, the user will pay one cent, but if the
price is one dollar, they will pay one dollar.  If it goes to one
dollar and one cent, then the instance vanishes, along with all of the
data on it.  This requires some care and thought with the bidding, as
the instance price can fluctuate significantly, as shown in the
example in Figure~\ref{spot_price_example}.

\subsection{Costs}

There are several factors to consider when deciding between using
  cloud based services or purchasing hardware.  One of the most
  important of these for many users is the cost.  One must consider
  whether it will cost more to develop and use an analysis pipeline on
  a cloud server or on a private compute cluster.  A few of issues we
  found important for estimating the cost different were the duration
  of the project, the computing demands profile, and the likelihood of
  other projects that may also be able to run on the cloud in the
  future.

Project duration is important to consider when estimating the
  costs of the cloud vs. purchasing hardware.  Both short and long
  projects may suffer from purchasing hardware.  Short projects may be
  delayed waiting for hardware to be ordered, delivered, set up, and
  managed.  Long projects may develop hardware that runs past its
  service contract, fails, or becomes obsolete.  In general, projects
  of about 3 years tend to benefit from hardware purchase, since that
  uses the full typical service contract window, and if getting going
  takes a couple of months, that is only a small fraction of the
  project time.

The computing demands profile is also important to consider.
  Even if the project is an appropriate duration for a typical
  hardware contract, the project may only need a large amount of
  processing power for a few weeks per year.  In this case, 100
  instances on a cloud system would only cost $\sim$\$5k per year, and
  a compute cluster may cost closer to $\sim$\$100k for the hardware
  alone.

It is also important to consider if the development costs to run
  on the cloud may pay off with future projects.  Perhaps the project
  has a duration and demand profile that put purchasing hardware and
  using cloud resources at similar cost.  In this case, perhaps
  developing the ability to use the cloud is worthwhile if it will
  allow future projects to use the ability.  For example, while the
  cost to develop the PHAT pipeline on EC2 was comparable to the cost
  to purchase our own cluster, we have been able to use the
  infrastructure we developed for many other {\it HST} projects.  On
  the other hand, if we had purchased hardware, it would be well past
  its expected lifetime by now.

Ideally it would be helpful to know how much it costs to develop
  a pipeline server that can be used in the way we outline in this
  paper.  However, because our pipeline was developed over the course
  of over a year, by several scientists and software engineers, and
  overcame many issues that no longer exist, it is difficult to
  estimate how much it would cost a team to do this today.  Still, it
  is likely to cost about 1 full-time employee for one year with
  significant experience with designing database systems and working
  with application programming interfaces (APIs) to build the master
  instance image, the job handler, and the monitoring system.  This
  estimate is likely an upper limit given the increased availability
  of tools to make this kind of computing easier to implement (see
  Appendix).  For us, this work was split as partial support for
  several people.  With overheads, this effort amounts to about
  $\sim$\$200k, which does not include the cost of the cloud computing
  itself (typically $\sim$\$0.05 per CPU hour for spot instances).

Developing a pipeline to run on purchased hardware has a cost as
  well.  In this case, one still needs to develop a job handling
  system and a way to monitor the jobs.  However, this kind of set up
  is much simplified because of the lower security concerns, shared
  storage, and scheduler software that come with most compute
  clusters.  In addition, one must pay for the purchase, power,
  storage, and administration of the hardware.  Typical hardware for a
  $\sim$20 node compute cluster would be $\sim$\$100k; adding in
  typical service contracts, storage, software licenses, and other
  expenses, brings this to $\gap$\$200k for 3 years of use.  Using
  such a cluster efficiently still requires some kind of pipeline
  architecture that, while simpler than developing a cloud-based
  pipeline, has its own development costs associated with it. All of
  these costs should be compared to the number of CPU hours one
  intends to use and the need for longer term applications to see how
  it compares to a one-time $\sim$\$200k development cost plus
  $\sim$\$0.05 per CPU hour, that one may expect to pay to do the
  project with cloud services.

\subsection{Machine Images}

Cloud computing systems rely on the user to supply a computing
operating system and environment that can be easily replicated on as
many machines as are requested. Most commercial cloud computing
services will likely use something like a full snapshot from a working
compute node to allow such replication.  Such snapshots provide users
with the flexibility to tailor their operating system to their needs.
They contain a complete copy of the software and data needed to start
an individual worker machine, including the operating system.  In case
of AWS, these snapshot files are called Amazon Machine Images (AMIs).

Amazon has several options for starter AMIs, which contain enough
capabilities to allow an instance to boot and allow a user to log in.
Then, the user can customize the AMI by installing the software they
require for their data analysis.  Once all of that software is
installed and working properly, the user can save the AMI under their
own account.  Then, the user can replicate as many copies of that AMI
as often as they wish.  Thus, once the user has an AMI capable of
running their data analysis software, they can generate 1, 10, or
thousands of instances with those capabilities.

\subsection{Storage}

One challenging aspect to analyzing data on the cloud is the lack of
shared disk space between machines.  In most cloud systems, the
individual machines only see their own local storage, not that of
other machines in the cloud.  In our AWS example, instances on EC2 act
as stand-alone computers. Their storage is local, and disappears when
the user turns off the instance.  However, Amazon has a very large
separate storage service, called Simple Cloud Storage Service‎ (S3),
which is always available to any machine connected to the internet.
Any data pushed from the instance to S3 will be safe.  Thus, a
significant challenge in leveraging many instances is making sure that
they each have the correct version of any file that is being analyzed.

\begin{figure*}
\includegraphics[width=7.0in]{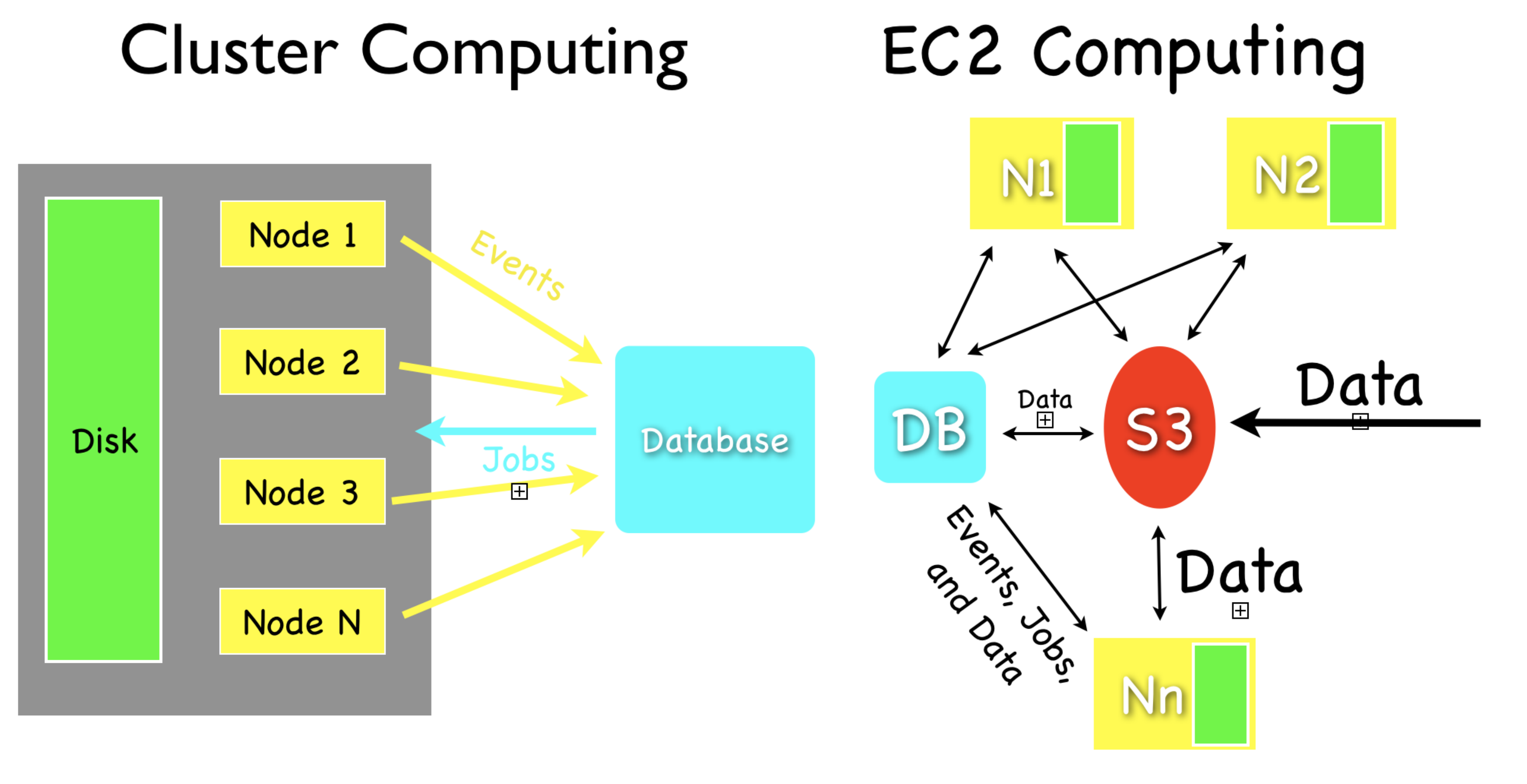}
\caption{{\it Left:} Schematic of computing on a traditional computing cluster, where all compute nodes share disks and the system has a built-in job scheduler. {\it Right:} Schematic of how one replicates a similar system on a computing cloud, where data must be shared between nodes through a common storage system that is not mounted directly to the machines.}
\label{computing_schematic}
\end{figure*}

We show the storage design in Figure~\ref{computing_schematic}.  For
example, if a process updates the cosmic ray masking of an image, and
another process runs searches for celestial light sources on the
image, then the instance that runs the former must push its updated
file to S3, and the instance that runs the latter must pull that
updated version from S3.  Keeping files synchronized across the worker
nodes proved to be very challenging and required a significant amount
of custom software when we built our pipeline.  However, the advantage
of all of this effort is that all of the most up-to-date products are
always on S3, which simplifies recovery from losing instances and
makes all products instantly available anywhere in the world.

\subsection{Security}

To use EC2, or likely any commercial cloud service, one must have an
account which gets charged for the resources used. On AWS, this
account comes with keys that allow one to start instances and move
data.  The charge for the resources is then applied to the account
associated with the keys.  For large amounts of data reduction, it is
critical to be able to quickly start, track, and terminate many
instances.  For us, this meant automating spot requests for new
instances and the ability to for those instances pull data from and
push data to S3.  Thus our instances have our keys, and if they were
successfully hacked, the hackers could take our keys and run a huge
amount of computing, billing it to our accounts without us ever
finding them.  Therefore, security is extremely important.

A surprisingly large amount of the development to move to the
commercial cloud was automating all of the security calls in a safe
way.  Two key parts to this were making sure there was only one open
port on our nodes, and only allowing information with very specific
formats through those ports.  Limiting the communication possible with
an instance is straight-forward in EC2, but making it work with an
automated pipeline architecture is a challenge, which we will describe
further in Section~\ref{architecture}.

\section{Pipeline Architecture}\label{architecture}

A software pipeline is typically understood to be a series of scripts
run on a set of input files that produces useful products from those
files.  In the case of PHAT, the pipeline started from the
HST-provided calibrated images, and from those images, produced clean
mosaic images, in all bands, color images, catalogs of resolved
stellar photometry, artificial star tests, and many diagnostic and
scientific plots from each step in the process.

\begin{figure*}
\begin{center}
\includegraphics[width=6.0in]{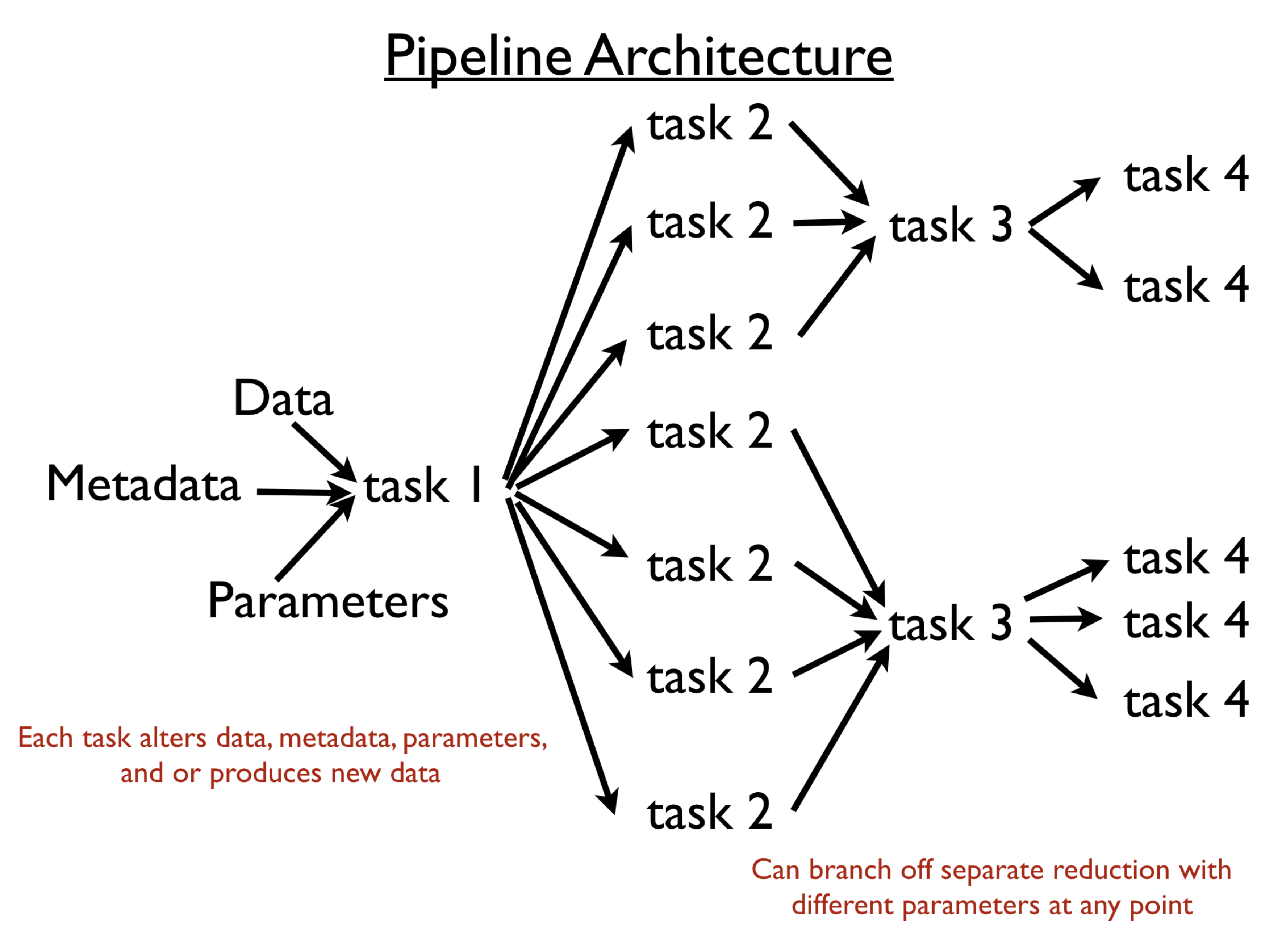}
\end{center}
\caption{Schematic of our pipeline architecture, where events start tasks, which then produce more events to complete the analysis.}
\label{pipeline_schematic}
\end{figure*}

To keep track of all of these tasks while they were being performed on
thousands of images, we adopted a master-worker pipeline architecture,
which we show with a schematic drawing in Figure~\ref{pipeline_schematic}.
This architecture requires one machine that keeps track of all the
jobs being done while any number of other machines actually do the
jobs.  The master machine needs to be a server that is able to
communicate with many other machines simultaneously.  It does not
require any analysis software, but instead requires sophisticated
communication and security software.  The worker machines can all be
identical as long as they all have the appropriate software to perform
the analysis jobs and to report back to the master machine.

To run a master-worker pipeline requires two AMIs: one for the master,
and one for the workers.  Our master AMI is a specially configured
(networking, security, backup, SVN-repository) Linux-box (AWS Linux or
Ubuntu) functioning as a webserver (Apache$+$Passenger or
Unicorn$+$Nginx) running a highly customized web-application (Ruby on
Rails) interfaced with a traditional database (mySQL) that is queried
by scripts running on worker instances via HTTP requests which are
generated and the corresponding responses parsed using an utility
suite (Java).  We ran this AMI on an on-demand instance, so that we
could be sure the server would stay up.  More efficient may be to host
the server on a reserved instance if the user knows they'll be using
their pipeline semi-regularly for a long time, whereas a more economic
solution is running the server on a spot instance with an unusually
high bid and a reliable, frequent backup system if the pipeline
infrastructure will only be used intermittently.

For reducing the PHAT software, our worker AMI needed many custom
additions, including PyRAF, IDL, DOLPHOT, and all of our pipeline
communication software, which in our case was Perl scripts that called
our suite of Java tools for querying the server.  These tools had all
of the necessary capabilities back in 2009 when we started the
project.  For any future pipeline builds, we would recommend using
python since it has all of the necessary capabilities now and is
better supported by the community.  Also, keeping all of the steps
within one language will make finding and fixing bugs much more
straight-forward.

One advantage of the master-worker architecture for cloud computing is
that one can request as many worker nodes as one needs to get the work
done, and then scale back when the work completes.  Furthermore, if
the server is designed to recover from failures efficiently, all of
the worker nodes can be on spot instances, reducing the compute cost
by up to an order of magnitude.  We set up our pipeline for reducing
the PHAT data using spot instance workers.

Our pipeline used an event based routine to keep track of jobs.  Each
job consists of an executable script called a task, the conditions
under which the task is to be run, and any options that need to be
passed to that task.  Before running data through the pipeline, the
user registers all of the executable scripts that are part of the
pipeline.  This enters the names and options for those scripts into a
database that records how events trigger the tasks.  A task may
produce events that trigger many other tasks.  These can then produce
their own events that trigger even more tasks, and so on, until the
pipeline is complete.  Each event is stored as an entry in the
database, and it contains all of the information needed to run the
task that it triggers.  Thus, if a task fails, it can easily be
restarted simply by pointing the pipeline to the event that triggered
that task.  In this way, recovering from failed tasks is simple and
efficient, as one never needs to rerun successfully completed tasks.

This architecture is especially important for EC2, since a spot
instance worker can disappear at any time, causing all of its running
tasks to fail.  In addition, tasks may fail due to some other problem,
such as a data set that has different properties that the pipeline
task expects or a failed data transfer from S3.  Virtually all
failures are efficient to fix within our architecture, which also
allows us to debug a task and then retry it without restarting the
pipeline run from the beginning.

Furthermore, this architecture makes it possible to create an
extremely secure pipeline by pushing all communication through a
single instance.  The server is the largest security risk, as it must
be allowed to receive data; however, it can be configured to only
allow incoming data of a format known only to the worker nodes, and
the worker nodes can be configured to only accept incoming data from
the server's ip address.  Since all of the workers are copies of one
another, this security can be part of the worker AMI.  This technique
makes the user free to carefully set up and monitor the security of
only one instance to keep all of the accounts secure.

\section{Monitoring} 

\begin{figure*}
\includegraphics[width=7.0in]{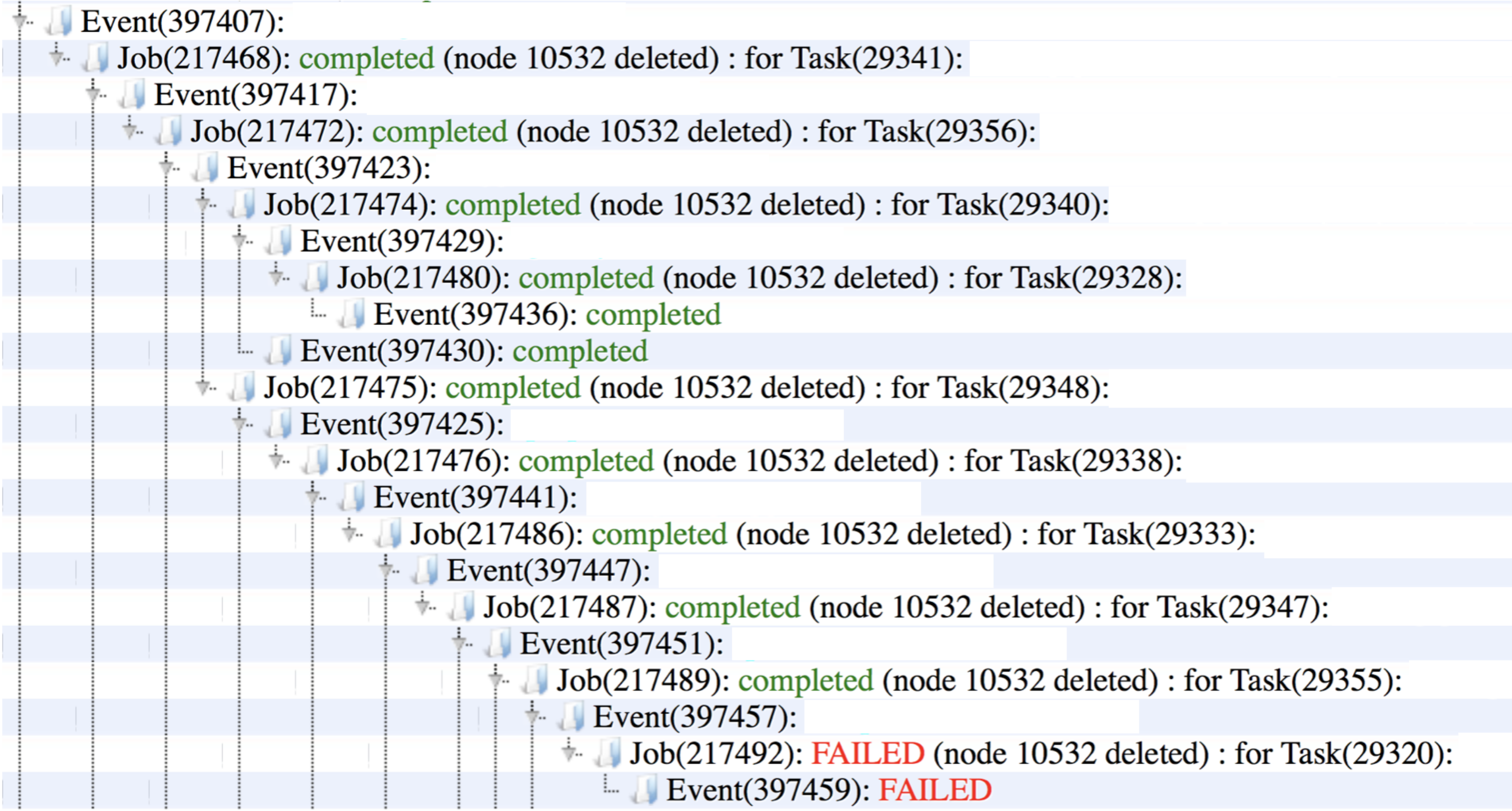}
\caption{Example of a pipeline monitoring page.  This kind of display makes is simple to find failed jobs, such as 217492 in this example.  Then the user can quickly find and investigate problems.}
\label{job_tree_example}
\end{figure*}

Because the pipeline performs many tasks on a large number of data
sets, it is critical to have a simple way to monitor the events and
jobs to easily spot, investigate, or restart failures.  Because our
pipeline architecture works by keeping a database of tasks and events,
we are able to simply display these tasks and events, along with their
pipeline state and the instance on which they were executed, in a tree
diagram.  This tree is populated directly from the database, and it
simply shows all of the work that has been and is currently being done
within the pipeline, allowing an easy scan by eye or with a find tool,
to look for any problems.  When a problem is spotted, it is then easy
to find the data associated with the problem through the information
associated with the event.  We show an example from our system in
Figure~\ref{job_tree_example}, where we included a failed job to
highlight how this system makes it easy to track and troubleshoot such
problems.

\section{Photometry Database}

Once the computing is complete, a huge amount of reduced data are
available.  These reduced data are in the form of tables that have
rows of measurements.  Each measurement typically will have many
values associated with it, such as the time the observation was taken,
the filter, the count rate measured, the background level, and any
quality metrics about the pixels measured.  Putting all of this data
into an efficient and intuitive database is the next challenge in the
process of managing such large observing programs.

For PHAT, our original data release simply supplied the images and
catalogs in FITS format.  While these tables
and images contain reliable measurements and all of the necessary
meta-data for any user to find an measurement available within the
data, they are numerous, large, and complex, forcing any user to learn
quite a bit about data structures in order to write code to find what
they are looking for.  This method for data distribution therefore
does not lend itself to simple use for users interested in just a
small portion of the data, such as a single object of interest.

The NOAO Data Lab \citep{Fitzpatrick2016} became publicly available
in June 2017.  The goal of the Data Lab is to enable efficient
exploration and analysis of large datasets, including catalogs. Among
its features are a database for catalog data, accessible both through
a Table Access Protocol (TAP) service and through direct PostgreSQL
queries, web-based and programmatic catalog query interfaces, remote
storage space for personal database tables and files, a
JupyterHub-based notebook analysis environment, and a Simple Image
Access (SIA) service.  

We had a few reasons for hosting the PHAT catalog data in the NOAO
Data Lab.  First, it is being developed as a site for long-term
hosting of major surveys such as the Dark Energy Survey, the Legacy
Survey (the DESI targeting survey), and others, and thus ensures that
the PHAT catalog will have a long shelf life.  Second, the Data Lab’s
existing tools and planned improvements (such as machine learning
tools), will add significant analysis capability for users of the PHAT
catalog now and in the future.  Finally, while the current analysis
model is based on notebooks running in JupyterHub, in the future Data
Lab will allow fully scripted workflows running in Containers on its
hardware, thus allowing for efficient automated analysis of PHAT and
other datasets.

The Data Lab has some limitations on the resources provided to users,
but in many cases these are soft and can be increased on a per-user
basis as needed.  For synchronous database queries, the standard time
limit is 600 s, while for asynchronous queries, it is 12 hours.  These
limits should accommodate the vast majority of PHAT catalog use cases.
For storage, the per-user limit is 1 TB for VOSpace virtual storage,
while for myDB it is 100 GB, both of which can accommodate a
significant fraction of the full PHAT catalog.  For analysis in a
Jupyter notebook, limits are imposed by the shared memory of the host
server, which for practical purposes can accommodate ~10 GB per user,
or $\sim$25\% of the full PHAT object catalog.

For PHAT, we ingested the combined photometric measurements for all
objects in all bands and the measurements from the individual
exposures, along with their meta-data, into tables in the Data Lab
database.  These tables allow fast access to the PHAT photometric
data, and allow users to to produce sparse lightcurves and potentially
look for proper motions.

Although there are not many epochs in the PHAT photometry, the data do
contain a large amount of variability data.  The observations were
performed in 3 hour visits over four years.  Each visit contained
several individual exposures, allowing searches for variability on
3-hour timescales survey-wide.  Furthermore, because the survey was
tiled with the WFC3/IR camera, the WFC3/UVIS and ACS/WFC channels
contain overlapping measurements for multiple visits over portions of
the survey.  This overlap allows comparisons across multiple visits,
which can be separated by hours up to years, making it sensitive to
variability on these timescales over portions of the survey.  The
available timescales change drastically with position, but in
principle these overlaps can be used to probe longer timescales of
variability.  Finally, because the photometry for each visit was
performed separately, there are separate measurements of the position
of the sources in overlapping visits.  These positions can, in
principle, be compared to look for proper motions.  The original data
release, which is simple catalog files, makes such studies possible,
but difficult.

To provide the necessary search capabilities to tap into the
variability and proper motions information in the PHAT data, we have
developed a schema for the database that allows each source to have a
global identifier, combined measurement for each visit in each band
that covered the sources in that visit, as well as individual
measurements from each individual exposure taken during that visit.
Along with the observation time, each measurement also contains all of
the data quality information returned by DOLPHOT.

\begin{figure*}
\includegraphics[width=7.0in]{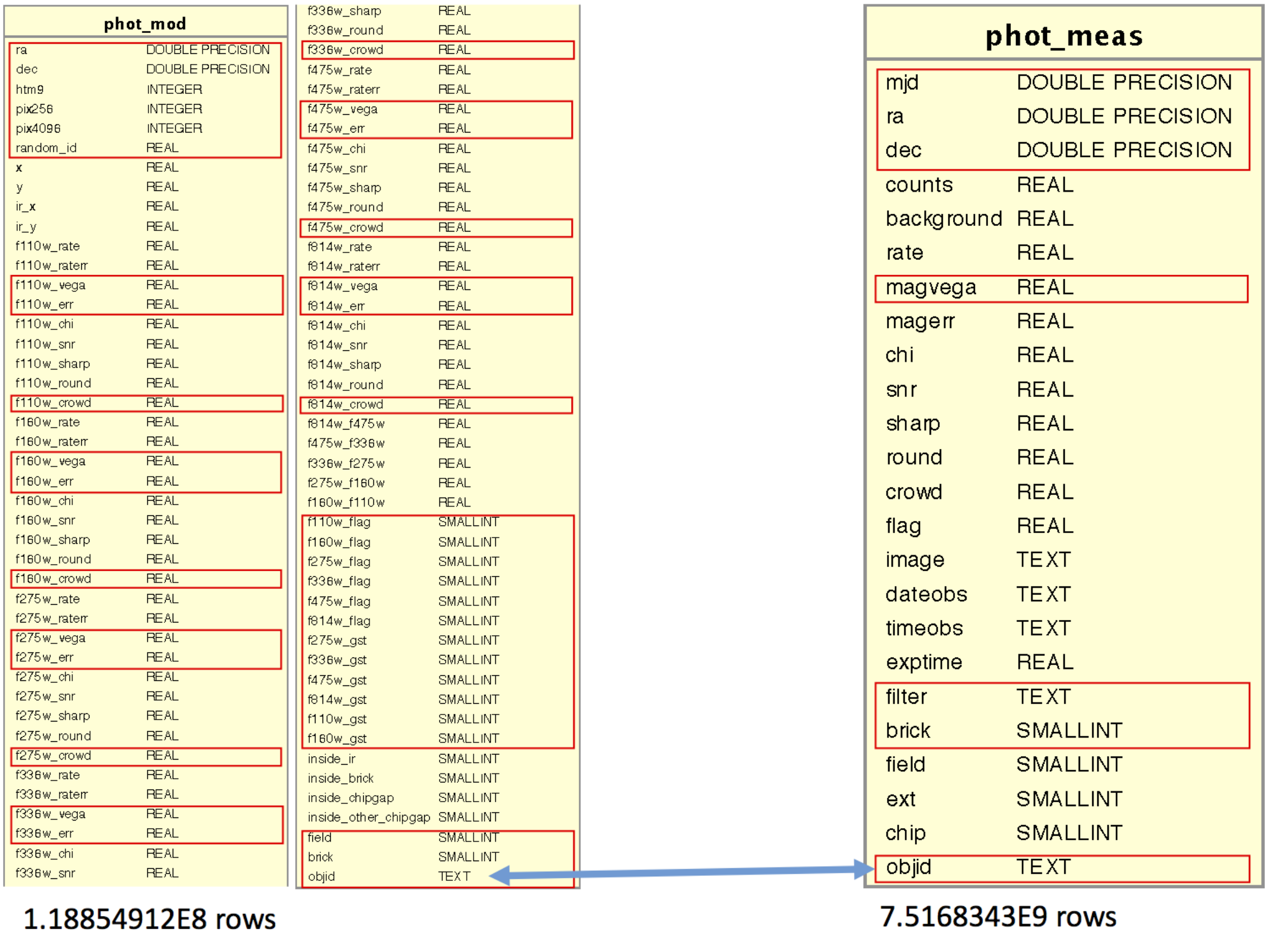}
\caption{Entity relationship diagram for the PHAT database tables.  The {\tt phot\_mod} table contains the average photometric measurements per object, with 88 columns of measurements and $>$118 million rows, one object per row.  The {\tt phot\_meas} table contains all of the single epoch measurements for all of the objects, with 24 columns and $>$7.5 billion rows, one measurement per row.  Columns that contain indices in the database are outlined by red boxes, while the tables may be joined on the {\tt objid} column, the primary key of the tables.}
\label{table_schema}
\end{figure*}

The photometry is organized into two main tables within a database
called {\tt phat\_v2}; see Figure~\ref{table_schema} for the table
schema.  The table {\tt phot\_mod} ($\sim$118 million rows) contains a
unique object ID ({\tt objid}) as the primary key, a unique random
integer ({\tt random\_id} to allow for selection of random samples
from the table, the averaged positions, photometry, and data quality
flags for all of the objects, and spatial indices computed on the 9th
order Hierarchical Triangular Mesh (HTM) and HEALPix (both NSIDE=256
RING and NSIDE=4096 NEST) systems.  Many of the columns have database
indices computed to allow for fast constraints on values in those
columns.  The data is clustered on the RA and Dec and spatially
indexed with Q3C \citep{Koposov2006} to allow for fast cone
searches and spatial table joins.

The table {\tt phot\_meas} contains the photometry from the individual
epochs, with one row per measurement ($\sim$7.5 billion rows).  It
contains the same unique object IDs as the {\tt phot\_mod} table, the
times of the measurements expressed as modified Julian dates (MJD),
the measured photometry and photometric flags, and metadata such as
the filter used in the measurement, the exposure time, and the
filename of the image from which the measurement derived.  There are
database indices for the {\tt MJD, RA, DEC, MAGVEGA, FILTER, BRICK,
 and OBJID} columns, as well as a Q3C spatial index.

For completeness, the database also contains tables named {\tt
  phot\_v2}, which contain the table data in the same form as in the
FITS files produced by the PHAT pipeline.

\section{Data Lab Services}

As mentioned above, the NOAO Data Lab provides a number of data access
and analysis services for the PHAT dataset.  These include:

\begin{enumerate}

\item Anonymous and authenticated access to Data Lab services.
  Anonymous users can query the database and run temporary Jupyter
  notebooks on the public notebook server, but cannot store results in
  personal database storage (myDB), access the virtual storage system
  (VOSpace), or save Jupyter notebooks permanently.  By creating and
  logging into an account, authenticated users get access to a
  dedicated Jupyter notebook server with permanently stored notebooks,
  and 1 TB of storage space for personal database tables and files.  A
  Python authClient package is available for authentication within a
  Python script or notebook.

\item A web query form and schema browser.  The webpage
  http://datalab.noao.edu/query.php provides a form interface to all
  of the datasets available through the Data Lab, including the PHAT
  catalogs.  Clicking on the {\tt phat\_v2} database brings up the
  list of available tables, while clicking on a particular table
  brings up a list of the columns and a query interface.  The query
  interface allows ADQL \citep{Plante2007} queries with results
  returned to the browser or, for authenticated users, as tables in
  myDB or as files in virtual storage.

\item Queries through TOPCAT \citep{Taylor2005}.  By pointing TOPCAT at the
  URL of the Data Lab TAP service (http://datalab.noao.edu/tap), the
  PHAT catalog can be queried through ADQL and the results analyzed in
  the TOPCAT environment.

\item Python and command-line query clients.  The Data Lab client
  package (https://github.com/noao-datalab/datalab-client) contains
  the multipurpose {\tt datalab} command-line interface and the
  queryClient Python module.  Both interfaces allow both synchronous
  (for which control is suspended until a result is returned) and
  asynchronous (for which the query operation is given a job ID and
  run in the background) queries, using both ADQL and standard SQL
  syntax.  The queryClient module is pre-installed on the Data Lab
  Jupyter notebook server.

\item An image cutout service.  The Data Lab SIA service provides
  access to cutouts of the PHAT Drizzled images.  For a given position
  on the sky, the SIA service returns a table of metadata of all
  images that fall within a specified radius.  The metadata includes
  select header information for each image, along with a URL to
  retrieve a cutout of a specified size.

\item Two JupyterHub notebook servers, one for temporary anonymous
  use, and the other for long-term authenticated use.  These servers
  provide access to common Python libraries as well as all Data Lab
  Python modules, including the authClient authorization module, the
  queryClient query module, the storeClient virtual storage module,
  and other interface modules.  The Jupyter notebook servers provide a
  convenient way to run code close to the data.

\end{enumerate}

More information on using these services is available on the Data Lab
web page, http://datalab.noao.edu.

\section{Example Case: Basic Queries, Maps, and Sparse Lightcurves}

\begin{figure*}
\includegraphics[width=7.0in]{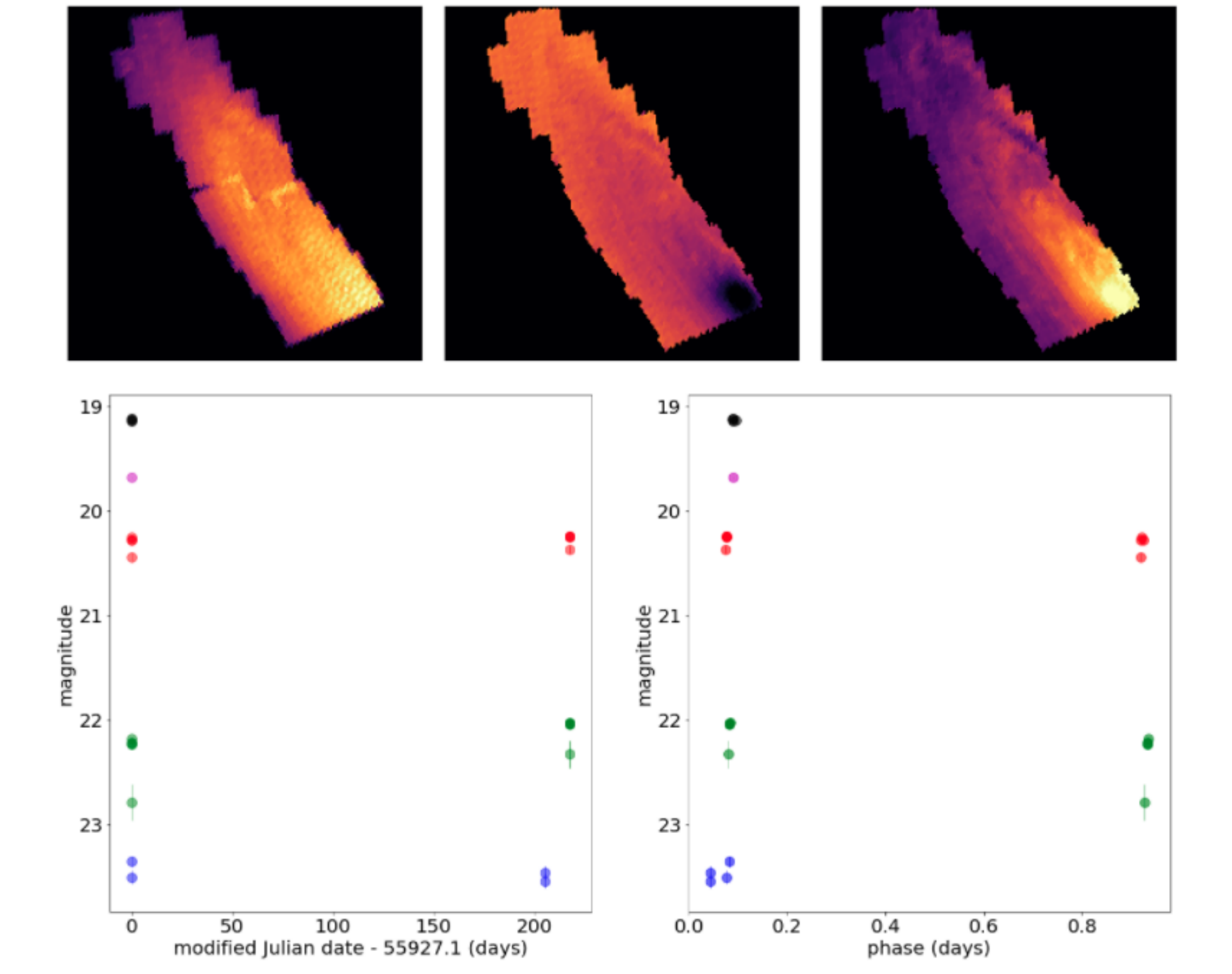}
\caption{Figures from our PHAT example Jupyter notebook.  Top row: maps of the object density (left), mean magnitude (middle) and mean color (right) obtained by issuing queries to the {\tt phat\_mod} table and aggregating by HEALPix (NSIDE=4096) index.  The queries run quickly, allowing efficient exploration of mean quantities of all 118 million objects.  Bottom row: the time series (left) and period-folded light curve (right) of the Cepheid variable V5343 \citep{Wagner-Kaiser2015} from a query of the single epoch {\tt phot\_meas} table.}
\label{notebook_results} 
\end{figure*}

As an example of analysis of the PHAT dataset, we have created a
Jupyter notebook to demonstrate simple queries of the PHAT catalog,
how to produce maps of quantities such as object density, mean
magnitude, and mean color from the catalog, and how to create sparse
light curves from the single epoch catalog.  Figure
~\ref{notebook_results} shows some of the output from the notebook;
the full notebook is available at http://datalab.noao.edu.  A simple
query is to retrieve the first 100 rows from the object catalog.  The
query is written as a string in Python, and then called with the
queryClient:\\

{\tt
\noindent
token = authClient.login('anonymous')\\
query = "SELECT * FROM phat\_v2.phot\_mod LIMIT 100"\\
result = queryClient.query(token,sql=query)\\
}

To make maps of aggregate quantities like number density, mean
magnitude, or mean color, we use the database to group results by
HEALPix number:\\

{\tt
\noindent
query = "SELECT avg(ra) as ra0,avg(dec) as dec0,pix4096,count(pix4096) as nb FROM phat\_v2.phot\_mod GROUP BY pix4096"\\
}

\noindent
where {\tt pix4096} is the HEALPix number with NSIDE=4096 (NEST scheme).  Queries of this form were used to produce the maps in Figure ~\ref{notebook_results}.

In order to retrieve time series measurements of specific objects, we
need to query the {\tt phot\_meas} table.  For example, to retrieve
the PHAT time series \citep{Wagner-Kaiser2015} of DIRECT Cepheid V5343 \citep{Kaluzny1998,Stanek1998,Kaluzny1999,Stanek1999,Bonanos2003}, we perform a cone search within 0\farcs2 of its
position:\\

{\tt
\noindent
query="SELECT * FROM phat\_v2.phot\_meas WHERE q3c\_radial\_query(ra,dec,11.02203,41.23451,0.2/3600)"\\
}

\noindent
which returns all of the measurements for objects found.  The sparse time series and period-folded light curve are shown in Figure ~\ref{notebook_results}.


\section{Conclusions}

We have described some of the problems and solutions to reducing large
observational data sets on commercial compute cloud resources.  These
include the production of secure and efficient machine images, a
flexible pipeline architecture, and a comprehensive monitoring system.
With these pieces well-engineered it is possible to make data
processing on the cloud cost effective, as we have done for the PHAT
survey.

Furthermore, we have discussed the serving of the output data products
in a database that includes the quality metrics for each measurement
as well as the time.  While this system is not new for serving data,
as it was the main release method for SDSS, we have now produced such
a database for the PHAT photometry, for which we have described a
sample use case.

Support for this work was provided by NASA through grant GO-12055 from
the Space Telescope Science Institute, which is operated by the
Association of Universities for Research in Astronomy, Incorporated,
under NASA contract NAS5-26555. We also thank Amazon Research Grants
for their support in providing some of the compute time necessary for
the PHAT project.  Finally, we thank Julianne Dalcanton for her
excellent leadership of the PHAT project.


\clearpage

\clearpage

\section{Appendix:  Updated Tools for Cloud Computing}

Our photometry pipeline infrastructure has proven to be a highly
productive and durable science enabling tool over the past
decade. However, it highlights features available since $\sim$2009.
Our system is now nearly 10 years old, and there are more services
available to simplify the creation of such a pipeline.  For example,
AWS features now include automatic scaling of worker instance numbers,
fault tolerance (recovery, re-run of failed jobs) and computational
load balancing across a fleet of spot instances.  If we were to build
a master-worker pipeline today, we would likely use new tools made
available by the cloud service providers.  Here we discuss some of the
newer AWS applications.

For some purposes, off-the-shelf cloud computing solutions from
commercial vendors can replicate the capabilities of our pipeline with
less development effort.  For example, one can now use AWS\,Batch (and
similar services from other cloud services).  This service requires
the users to upload raw data, analysis scripts and compiled binaries
to AWS\,S3 storage along with a AWS\,Cloud\,Formation (JSON format)
configuration file describing their computational requirements. The
processing is then efficiently executed using the minimum amount of
resources needed only for the duration that they are needed. This
model is similar to batch processing on clusters or supercomputers and
is more easily adopted to different platforms. For comparison,
utilization of an AWS\,Elastic\,Map\,Reduce (EMR) cluster can utilize
more efficient architecture and applications available on the Hadoop
platform (e.g., Presto / Hive QL instead of SQL, Apache Spark for
distributed processing, Apache Hbase for storage etc.) but would
require the analysis pipeline to be customized for this platform.

Furthermore, there are now services available that essentially mimic
much of the functionality of our pipeline without the need to develop
them from scratch.  The master webserver itself would be replaced by
either AWS\,Lightsail (easily scalable) or AWS\,Elastic\,Beanstalk
(more economic), with the pipeline database (jobs, tasks, events etc.)
residing on AWS\,Relational\,Database\,Service (RDS).  Such a change
would remove the need for an on-demand server instance to always be
running.

The analysis scripts and compilable code could be managed under
AWS\,CodeCommit version control.  These would then be deployed for
each worker using AWS\,CodeBuild on a fleet of AWS\,Lambda (for
small,quick task) and AWS\,EC2 (for processor intensive tasks such as
running DOLPHOT) instances accessing AWS\,Elastic\,Block\,Storage
(EBS, only accessible by individual instances) or
Elastic\,File\,System (EFS, networked file systems can accessible by
multiple instance or remotely). Raw and processed images along with
photometry tables can be marshaled to/from STScI archives and
AWS\,S3/ECS/EFS disks by the AWS\,Data\,Pipeline. 

Potentially, even the data serving could be put on a cloud service.
For example, the photometry tables can be put into searchable
AWS\,Aurora databases using AWS\,Glue hosted on AWS\,Redshift that can
be queried using AWS,Athena and analyzed (e.g., for variability or
anomalies) using AWS\,Machine\,Learning (ML) and
AWS\,Artificial\,Intelligence (AI) tools.  However, this kind of
service would require a budget to host, unlike the free services
supplied by the NOAO Data Lab or the MAST High Level Science Products.

We would like to emphasize that our mention of various AWS offerings
by name is merely due to our greater familiarity with this particular
cloud service provider. Similar applications from other reputable
cloud vendors or through open-source applications (e.g., Openstack
Kubernetes) are likely to be just as suitable, but inventorying the
cloud marketplace is beyond the scope of this work. Before deciding to
use cloud resources for data analysis needs, it is important to
perform some cost benefit analysis to determine that it will be
beneficial in the long term.  We have found it beneficial when one
needs many instances occasionally over many years, since purchased
computers may sit idle, become obsolete, and lose their warranties
over these timescales.

\end{document}